\documentclass[prl,aps,twocolumn]{revtex4}

\usepackage{psfrag}
\usepackage{graphicx}
\usepackage{dcolumn}
\usepackage{latexsym,amsfonts}
\usepackage{bm}
\usepackage{amssymb}
\begin{document}

\title{Time Modulation of K--Shell Electron Capture Decay Rates of
H--Like Heavy Ions and Neutrino Masses}

\author{R. H\"ollwieser${^{\,a}}$, A. N. Ivanov${^{\,a}}$,
P. Kienle$^{b,c}$, M. Pitschmann${^{a,d}}$}
\affiliation{${^a}$Atominstitut der \"Osterreichischen
Universit\"aten, Technische Universit\"at Wien, Wiedner Hauptstrasse
8-10, A-1040 Wien, Austria} \affiliation{${^b}$Stefan Meyer Institut
f\"ur subatomare Physik \"Osterreichische Akademie der Wissenschaften,
Boltzmanngasse 3, A-1090, Wien, Austria}\affiliation{${^c}$Excellence
Cluster Universe Technische Universit\"at M\"unchen, D-85748 Garching,
Germany}\affiliation{$^d$University of Wisconsin--Madison, Department
of Physics, 1150 University Avenue, Madison, WI 53706,
USA}\email{ivanov@kph.tuwien.ac.at}

\date{\today}

\begin{abstract}
We calculate the neutrino masses, using the experimental data on the
periods of the time modulation of the K--shell electron capture $(EC$)
decay rates of the H--like heavy ions, measured at GSI.  The
corrections to neutrino masses, caused by interaction of massive
neutrinos with the strong Coulomb field of the daughter ions, are
taken into account.  
\end{abstract}

\maketitle

\subsection{Introduction}

Recently \cite{Ivanov3} we have analysed a process by which massive
neutrinos, produced in weak decays of the H--like heavy ions
\cite{GSI2}, can acquire non--trivial mass--corrections
caused by the interaction of massive neutrinos with a strong Coulomb
field of the daughter ions. We have shown that due to the dissociation
of massive neutrinos into $\ell^-W^+$ pairs in a strong Coulomb field
$\nu_j \to \sum_{\ell}U_{j\ell}\ell^-W^+$, where $\ell^-$ is a
negatively charged lepton electron $(e^-)$, muon $(\mu^-)$ or
$\tau$--lepton $(\tau^-)$ and $W^+$ is a $W$--boson, massive neutrinos
get corrections to their energy and mass, dependent on the relative
distance between the neutrinos and the daughter ions. According to
this hypothesis, massive antineutrinos, produced in the
continuous-state and bound-state $\beta^-$--decays of heavy ions,
should acquire mass--corrections with a sign opposite to the
mass--corrections for massive neutrinos due to the dissociation into
$\ell^+W^-$ pairs $\tilde{\nu}_j \to \sum_{\ell}U_{j\ell}\ell^+ W^-$,
where $\ell^+$ is a positron $(e^+)$ and positively charged $\mu^+$
and $\tau^+$.

Measurements of the K--shell electron capture ($EC$) decay rates of
the H--like ions ${^{142}}{\rm Pm}^{60+}$, ${^{140}}{\rm Pr}^{58+}$,
and ${^{122}}{\rm I}^{52+}$ (preliminary) at GSI in Darmstadt
\cite{GSI2} showed a time modulation of exponential decays with
periods $T_{EC} = 7.10(22)\,{\rm s}, 7.06(8)\,{\rm s}$ and
$6.11(3)\,{\rm s}$ and modulation amplitudes $a^{EC}_d = 0.23(4),
0.18(3)$ and $0.22(2)$ for ${^{142}}{\rm Pm}^{60+}$, ${^{140}}{\rm
Pr}^{58+}$ and ${^{122}}{\rm I}^{52+}$, respectively, defined by
\begin{eqnarray}\label{label1}
  \lambda^{EC}_d(t) = \lambda_{EC}\,(1 + a^{EC}_d \cos(\omega_{EC}t +
\phi_{EC})),
\end{eqnarray}
where $\lambda_{EC}$ is the $EC$--decay constant, $a^{EC}_d$, $T_{EC}
= 2\pi/\omega_{EC}$ and $\phi_{EC}$ are the amplitude, period and
phase of the time--dependent term \cite{GSI2}.
Furthermore it was shown that the $\beta^+$--decay rate of
${^{142}}{\rm Pm}^{60+}$, measured simultaneously with its modulated
$EC$--decay rate, is not modulated with an amplitude upper limit
$a_{\beta^+} < 0.03$ \cite{GSI2}.

As has been proposed in \cite{Ivanov2}--\cite{Ivanov4}, such a periodic
dependence of the $EC$--decay rate can be explained by the
interference of neutrino mass--eigenstates. The period of the time
modulation $T_{EC}$ has been obtained as
\begin{eqnarray}\label{label2}
T_{EC} = \frac{4\pi \gamma M_m}{(\Delta m^2_{21})_{\rm GSI}},
\end{eqnarray}
where $M_m$ is the mass of the mother ion, $\gamma = 1.432$ is the
Lorentz factor of the ions in the experimental storage ring
\cite{GSI2}.  According to \cite{Ivanov3}, $(\Delta
m^2_{21})_{\rm GSI}$ should be defined as follows
\begin{eqnarray}\label{label3}
(\Delta m^2_{21})_{\rm GSI} =(m_2 + \delta m_2)^2 - (m_1 +
 \delta m_1)^2,
\end{eqnarray}
where $m_2$ and $m_1$ are neutrino masses and $\delta m_2$ and $\delta
m_1$ are neutrino mass--corrections, caused by the interaction of
massive neutrinos with the strong Coulomb field of the daughter ion as
proposed in \cite{Ivanov3}.

From the experimental data on the periods of the time modulation of
the $EC$--decay rates, the masses $M_m \simeq 931.494\,A$ of mother
ions and Eq.(\ref{label2}) we get the following values for quadratic
mass difference $(\Delta m^2_{21})_{\rm GSI}$ of massive neutrinos
\begin{eqnarray}\label{label4}
\hspace{-0.1in}(\Delta m^2_{21})_{\rm GSI} =
\Bigg\{\begin{array}{r@{~,}l} 2.20(7)\times 10^{-4}\,{\rm eV^2} &
{^{142}}{\rm Pm}^{60+} \\ 2.18(3)\times 10^{-4}\,{\rm eV^2} &
{^{140}}{\rm Pr}^{58+}\\ 2.19(1)\times 10^{-4}\,{\rm eV^2} &
{^{122}}{\rm I}^{52+}.
\end{array}
\end{eqnarray}
 The values are equal within their error margins and yield a combined
squared neutrino mass difference $(\Delta m^2_{21})_{\rm GSI} = 2.19
\times 10^{-4}\,{\rm eV^2}$. This confirms the proportionality of the
period of time modulation to the mass number $A$ of the mother nucleus
$T_{EC} = \kappa\,A$, where $\kappa = 4\pi \gamma \hbar M_m/A (\Delta
m^2_{21})_{\rm GSI} = 0.050(4)\,{\rm s}$ \cite{Ivanov2}.

The value $(\Delta m^2_{21})_{\rm GSI} = 2.19\times 10^{-4}\,{\rm
eV^2}$ is 2.9 times larger than that reported by the KamLAND $(\Delta
m^2_{21})_{\rm KL} = 7.59(21)\times 10^{-5}\,{\rm eV^2}$ \cite{KL08}.
In \cite{Ivanov3} such a discrepancy has been proposed to be explained
by the neutrino mass--corrections in the strong Coulomb field of the
daughter ions. Assuming that the $(\Delta m^2_{21})_{\rm KL} =
7.59(21)\times 10^{-5}\,{\rm eV^2}$, deduced from antineutrino -
antineutrino oscillations $\tilde{\nu}_e \to \tilde{\nu}_e$ at
KamLAND, represents difference of squared proper neutrino masses
$(\Delta m^2_{21})_{\rm KL} = m^2_2 - m^2_1$, and taking into account
the GSI experimental value $(\Delta m^2_{21})_{\rm GSI} =
2.19\times 10^{-4}\,{\rm eV^2}$ one can estimate the magnitude of
neutrino masses $m_j \sim 0.11\,{\rm eV}$ \cite{Ivanov3}.

However, such an estimate of neutrino masses is only qualitative,
since nobody took into account the contribution of antineutrino
mass--corrections to antineutrino masses, produced in the
$\beta^-$--decays of fission fragments in nuclear reactors
\cite{Spectrum1}. Indeed, as has been pointed out by Nakajima {\it et
al.}  \cite{Spectrum2}, the energy spectrum of antineutrinos, produced
in the $\beta^-$--decays of fission fragments, has been analysed with
the probability of the electron antineutrino oscillations
$\tilde{\nu}_e \leftrightarrow \tilde{\nu}_e$, defined by
\begin{eqnarray}\label{label5}
P_{\tilde{\nu}_e \leftrightarrow \tilde{\nu}_e}(E_{\tilde{\nu}_e}) = 1
- \sin^2(2\theta_{12})\,\sin^2\Big(\frac{\Delta m^2_{21}L}{4
  E_{\tilde{\nu}_e}}\Big),
\end{eqnarray}
where $\theta_{12}$ is the mixing angle \cite{PDG10}, $\Delta m^2_{21}
= m^2_2 - m^2_1$ is given by the proper neutrino masses, $L$ is the
distance between the source and the detector of antineutrinos and
$E_{\tilde{\nu}_e}$ is the antineutrino energy.

However, one can show that the antineutrino mass--corrections, caused
by interactions of massive antineutrinos with the Coulomb field of the
daughter nuclei in the final state of $\beta^-$--decays of fission
fragments, change the probability of the electron antineutrino
oscillations $\tilde{\nu}_e \leftrightarrow \tilde{\nu}_e$ as follows
\begin{eqnarray}\label{label6}
\hspace{-0.3in}&&P_{\tilde{\nu}_e \leftrightarrow
\tilde{\nu}_e}(E_{\tilde{\nu}_e}) = 1 -
\frac{1}{2}\sin^2(2\theta_{12})\nonumber\\
\hspace{-0.3in}&&\times\,\Big\{1 - \cos\Big[\arctan\Big(\frac{ \delta
m^2_1 - \delta m^2_2 }{2
E_{\tilde{\nu}_e}\lambda_{\beta^-}}\Big)\Big]\nonumber\\
\hspace{-0.3in}&&\times\,\cos\Big[\frac{\Delta m^2_{21}L}{2
E_{\tilde{\nu}_e}} + \arctan\Big(\frac{\delta m^2_1 - \delta m^2_2}{2
E_{\tilde{\nu}_e}\lambda_{\beta^-}}\Big)\Big]\Big\},
\end{eqnarray}
where $\lambda_{\beta^-_c}$ is the decay rate of the $\beta^-$--decay
of the fission fragment and $\delta m^2_j = 2 m_j \delta m_j$ are
antineutrino mass--corrections.  For $\delta m_j \to 0$ the
probability Eq.(\ref{label6}) reduces to Eq.(\ref{label5}).
\begin{figure}[t]
\includegraphics[width = 0.70\linewidth]{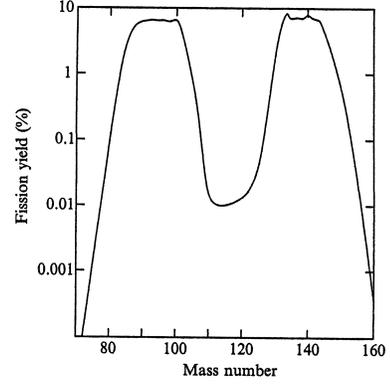}
\caption{Distribution of fission fragment masses from the fission of
${^{235}}{\rm U}$ induced by thermal neutrons \cite{Spectrum1}.}
\end{figure} The
probability Eq.(\ref{label6}) should be averaged over all fission
fragments unstable under $\beta^-$--decays. In thermal neutron fission
\cite{Spectrum1}, the fission fragments have a well known double hump
mass and nuclear charge $(Z)$ distribution (see Fig.\,1).

Thus, we can argue that the value $(\Delta m^2_{21})_{\rm KL} =
7.59(21)\times 10^{-5}\,{\rm eV^2}$, deduced from antineutrino -
antineutrino oscillations $\tilde{\nu}_e \to \tilde{\nu}_e$ by the
KamLAND experiment, does not correspond to the difference of squared
proper antineutrino masses $(\Delta m^2_{21})_{\rm KL} \neq m^2_2 -
m^2_1$. This means that neutrino mass--corrections should be used for
a quantitative explanation of the discrepancy between $(\Delta
m^2_{21})_{\rm GSI} = 2.19(3)\times 10^{-4}\,{\rm eV^2}$ and $(\Delta
m^2_{21})_{\rm KL} = 7.59(21)\times 10^{-5}\,{\rm eV^2}$.

In this letter we propose a model--independent calculation of neutrino
masses. We use the experimental data on the periods of the time
modulation of the $EC$--decay rates of the H--like like heavy ions
with different nuclear charge $Z$, measured at GSI, and corrections to
neutrino masses, caused by interactions of massive neutrinos with the
strong Coulomb fields of the daughter nuclei.

\subsection{Neutrino Masses}

For the calculation of neutrino masses we use the experimental data on
the time modulation of the $EC$--decay rates of a couple of ions
${^{A'}}{\rm X'}^{Z'+}$ and ${^{A''}}{\rm X''}^{Z''+}$.  The neutrino
masses we obtain as a solution of the system of two algebraical
equations
\begin{eqnarray}\label{label7}
\left\{\begin{array}{r@{\,=\,}l}
(m_2 + \delta m'_2)^2 - (m_1 + \delta m'_1)^2 & (\Delta m^2_{21})'_{\rm GSI}\\
(m_2 + \delta m''_2)^2 - (m_1 + \delta m''_1)^2 & (\Delta m^2_{21})''_{\rm GSI},
\end{array}\right.
\end{eqnarray}
where $\delta m'_j$ and $\delta m''_j$ are corrections to neutrino
masses in the Coulomb fields of the daughter ions of the $EC$--decays
of ${^{A'}}{\rm X'}^{(Z'- 1)+}$ and ${^{A''}}{\rm X''}^{(Z''- 1)+}$
ions, respectively. We can combine the GSI experimental data in three
systems of two algebraical equations Eq.(\ref{label7}). The solutions
of these systems of algebraical equations give neutrino masses.  The
Coulomb mass--corrections, required for the solution of the
Eq.(\ref{label7}) for every pair of the H--like heavy ions, are given
in Table I.
\begin{table}[h]
\begin{tabular}{|c|c|c|}
\hline ${^A}X^{(Z -1)+}$ & $10^4\,\delta m_1 (\rm eV) $ &
$10^4\,\delta m_2 (\rm eV) $ \\ \hline ${^{122}}{\rm I}^{52+} $ &
$-\,8.967 $ & $-\,5.068$ \\ \hline ${^{140}}{\rm Pr}^{58+}$ & $-\,10.680
$ & $-\,5.990$\\ \hline${^{142}}{\rm Pm}^{60+}$ & $-\,11.610
$ & $-\,6.507$\\ \hline
\end{tabular}
\caption{Numerical values for corrections to neutrino masses in the
strong Coulomb field of daughter nuclei, calculated for $R = 1.1\times
A^{1/3}$.}
\end{table}
\begin{table}[h]
\begin{tabular}{|c|c|c|c|}
\hline $(X', X'')$ & $({\rm I, Pr})$ & $({\rm I, Pm})$ & $({\rm Pr,
Pm})$\\ \hline $m_1\,(\rm eV)$ & $0.00657$ & $0.01345$ & $0.02982$\\
\hline $m_2\,(\rm eV)$ & $0.01636$ & $0.01991$ & $0.03292$\\ \hline
$m_3 \,(\rm eV)$ & $0.05165$ & $0.05288$& $0.05735$\\ \hline $\sum_j
m_j\,(\rm eV)$ & $0.07458$ & $0.08624$ & $0.12010$\\ \hline $10^4
(\Delta m^2_{21})_{\rm GSI}(X''')$ & $2.175$ & $2.196$ & $2.141$\\
\hline $10^4 \Delta m^2_{21}$ & $2.245$ & $2.155$ & $1.945$\\\hline
\end{tabular}
\caption{Neutrino masses, calculated for $R = 1.1\times A^{1/3}$
  \cite{Ivanov3}. The mass $m_3$ is calculated for $\Delta m^2_{32} =
  m^2_3 - m^2_2 = 2.40\times 10^{-3}\,{\rm eV^2}$ \cite{PDG10};
  $\Delta m^2_{21} = m^2_2 - m^2_1$.}
\end{table}
The neutrino masses are summarised in Table II.

In Table II the quantity $(\Delta m^2_{21})_{\rm GSI}(X''')$
corresponds to the $EC$--decay of the H--like heavy ${^{A'''}}{\rm
X'''}^{(Z'''- 1)+}$ ion, which we do not take into account in the
system of algebraical equations Eq.(\ref{label7}). In the last line of
Table II we calculate $\Delta m^2_{21} = m^2_2 - m^2_1$, where we use
the proper neutrino masses.

One can see that the neutrino masses, calculated from the experimental
data on the periods of the time modulation of the $EC$--decay rates of
${^{122}}{\rm I}^{52+}$ and ${^{142}}{\rm Pm}^{60+}$ ions, reproduce
better $(\Delta m^2_{21})_{\rm GSI} = 2.18(3)\times 10^{-4}\,{\rm
eV^2}$, calculated from the period of the time modulation of the
$EC$--decay rate of ${^{140}}{\rm Pr}^{58+}$ ions.  In addition the
neutrino masses are less sensitive to Coulomb mass--corrections.

As a result, the neutrino masses, given in the column $({\rm I, Pm})$
of Table II, can be recommended for the theoretical analysis of
neutrino reactions.

\subsection{Conclusion}

We have shown that taking into account the corrections to neutrino
masses, caused by the interactions of neutrino mass--eigenstates with
the Coulomb fields of the fission products, the probability of
electron antineutrino oscillations $\tilde{\nu}_e \leftrightarrow
\tilde {\nu}_e$ differs from the probability of the electron
antineutrino oscillations, calculated without Coulomb corrections to
the antineutrino masses. This implies that the experimental data
$(\Delta m^2_{21})_{\rm KL} = 7.59(21)\times 10^{-5}\,{\rm eV^2}$ by
the KamLAND \cite{KL08} cannot be used for the determination of
$\Delta m^2_{21}$ without applying Coulomb mass--corrections.

For the calculation of neutrino masses we have used the GSI
experimental data on the time modulation of the $EC$--decay rates of
the H--like heavy ions with different nuclear charges $Z$ and the
corrections to neutrino masses, caused by the Coulomb field of the
daughter nuclei in the $EC$--decays of the H--like heavy ions. The
calculated neutrino masses are given in Table II. They satisfy the
strict cosmological constraints for the sum of neutrino masses
$0.05\,{\rm eV/c^2} < \sum_j m_j < 0.17\,{\rm eV/c^2}$ \cite{JL06} and
the mass of the heaviest neutrino $0.02\,{\rm eV/c^2} < m_3 <
0.40\,{\rm eV/c^2}$ \cite{PDG10}. We notice that the neutrinos in our
analysis are massive Dirac particles with masses, obeying a direct
hierarchy $m_1 < m _2 < m_3$ \cite{PDG10}.

The values of the Coulomb corrections to neutrino masses depend on the
values of the mixing angles $\theta_{12}$ and $\theta_{23}$
\cite{Ivanov3}.  For the numerical analysis we have used $\theta_{12}
= 34^0$ and $\theta_{23} = 45^0$ \cite{PDG10} (see also
\cite{Ivanov3}). Since the mixing angles can be obtained from the
experimental data on the solar neutron fluxes \cite{PDG10}, for the
analysis of the solar neutrino data by SNO \cite{PDG10} instead of
$(\Delta m^2_{21})_{\rm KL} = 7.59\times 10^{-5}\,{\rm eV}^2$ we
propose to use the difference of squared proper neutrino masses
$\Delta m^2_{21} = 2.155\times 10^{-4}\,{\rm eV}^2$ (see Table II),
which is less sensitive to Coulomb corrections to neutrino masses, and
to find new mixing angles for massive neutrinos.

\end{document}